\documentclass[prd,aps,a4paper,superscriptaddress,twocolumn,nofootinbib]{revtex4}
\usepackage{graphicx}
\usepackage{color}
\usepackage{dcolumn}
\usepackage{bm}
\usepackage{slashed}
\usepackage{amsmath}
\usepackage{latexsym}
\usepackage{amssymb}
\usepackage{mathrsfs}
\usepackage{amsfonts}
\usepackage{url}
\allowdisplaybreaks
\begin{document}
\title{Accurate calculation of gravitational wave memory}

\author{Xiaolin Liu}
\affiliation{Department of Astronomy, Beijing Normal University,
Beijing 100875, China}
\author{Xiaokai He}
\affiliation{School of Mathematics and
Computational Science, Hunan First Normal University, Changsha
410205, China}
\author{Zhoujian Cao
\footnote{corresponding author}} \email[Zhoujian Cao: ]{zjcao@amt.ac.cn}
\affiliation{Department of Astronomy, Beijing Normal University,
Beijing 100875, China}

\begin{abstract}
Gravitational wave memory is an important prediction of general relativity. The detection of the gravitational wave memory can be used to test general relativity and to deduce the property of the gravitational wave source. Quantitative model is important for such detection and signal interpretation. Previous works on gravitational wave memory always use the energy flux of gravitational wave to calculate memory. Such relation between gravitational wave energy and memory has only been validated for post-Newtonian approximation. The result of numerical relativity about gravitational wave memory is not confident yet. Accurately calculating memory is highly demanded. Here we propose a new method to calculate the gravitational wave memory. This method is based on Bondi-Metzner-Sachs theory. Consequently our method does not need slow motion and weak field conditions for gravitational wave source. Our new method can accurately calculate memory if the non-memory waveform is known. As an example, we combine our method with matured numerical relativity result about non-memory waveform for binary black hole coalescence. We calculate the waveform for memory which can be used to aid memory detection and gravitational wave source understanding. Our calculation result confirms preliminary numerical relativity result about memory. We find out the dependence of the memory amplitude to the mass ratio and the spins of the two spin aligned black holes.
\end{abstract}

\maketitle

\section{Introduction}
The memory of gravitational wave (GW) was firstly found by Zeldovich, Braginsky, Thorne and their coworkers \cite{Zeldovich74,Pay83,Braginsky:1986ia,braginsky1987gravitational}. This memory effect is produced by the gravitational wave source directly. Later Christodoulou found that gravitational wave itself can also produce memory \cite{PhysRevLett.67.1486,Fra92}.

The memory found before Christodoulou is usually called ordinary memory. The ordinary memory is produced by the quadrupole moment change of the source. And the memory found by Christodoulou is called nonlinear memory. Thorne \cite{Tho92} assumed a relation between the gravitational wave flux and the nonlinear memory through analogy of `quadrupole moment change of gravitational wave energy'
\begin{align}
\Delta h_{jk}^{\rm TT}=\frac{4}{r}\int\frac{dE}{d\Omega'}\left(\frac{\xi^{'j}\xi^{'k}}{1-\cos\theta'}\right)^{\rm TT}d\Omega'.\label{assumeq}
\end{align}
This relation corresponds to the Eq.~(2) of \cite{Tho92}. The integral is over the solid angle $\Omega'$ surrounding the source, $E$ is the energy of gravitational wave, $\xi^{'j}$ is a unit vector pointing from the source toward $d\Omega'$, and $\theta'$ is the angle between $\xi^{'j}$ and the direction to the detector. The assumed relation (\ref{assumeq}) can be shown valid when the condition of post-Newtonian approximation is satisfied \cite{Tho92,WisWil91,BlaDam92}.

Recent years, many works including \cite{Fav09a,Fav09b,favata2009gravitational,favata2010gravitational,PhysRevD.84.124013,PhysRevD.95.084048,PhysRevD.98.064031,2020arXiv200906351K} applied the above assumed relation (\ref{assumeq}) to the full inspiral-merger-ringdown process of binary black hole to get the gravitational waveform of memory. And later these GW memory waveform was used to determine when LIGO would be able to detect the memory effect \cite{PhysRevLett.117.061102,PhysRevD.101.083026} and to search memory signal in LIGO data \cite{PhysRevD.101.023011,PhysRevD.101.104041}.

On the numerical relativity (NR) side, the calculation for non-memory waveform has become more and more accurate. The waveform extraction technique involved in NR guarantees the calculated gravitational wave is gauge invariant, which makes different numerical relativity groups using different Einstein equation formulation, different initial data form, and different coordinate condition during the evolution get the same waveform result (early references including \cite{Baker_2007}). The extracted waveform corresponds to the two polarization modes $h_+$ and $h_\times$. The reported gravitational wave events by LIGO and Virgo highly depend on gravitational waveform models including EOBNR, IMRPhenomena and surrogate models \cite{PhysRevLett.116.061102,PhysRevX.9.031040}. In contrast, numerical relativity results on memory are much less confident. Some preliminary NR results on memory have been got in \cite{PolRei11,PhysRevD.102.104007,2020arXiv201101309M}.

Theoretical model is very important to memory detection and signal interpretation \cite{Set09,VanLev10,PshBasPos10,CorJen12,MadCorCha14,Arzoumanian_2015,PhysRevLett.118.181103,Divakarla:2019zjj}. In this paper, we propose a new method to calculate the gravitational wave memory. This method is based on the Bondi-Metzner-Sachs (BMS) theory \cite{BonVanMet62,Sac62,PenRin88} in stead of the assumption (\ref{assumeq}). Since BMS theory does not need the conditions of slow motion and weak field for the GW source, this new method is very accurate for GW memory calculation. We adopt geometric units with $c=G=1$ through this paper.

\section{New method to calculate the gravitational wave memory}
Based on the Bondi-Metzner-Sachs (BMS) theory, gravitational radiation can be described at null infinity with Bondi-Sachs (BS) coordinate $(u,r,\theta,\phi)$. Here $u$ is called Bondi time which corresponds to the time of observer very far away from the GW source, say the GW detector. Inside the spacetime of the gravitational wave source which is looked as an isolated spacetime, the slice of constant $u$ is null. On the null infinity, the gravitational waveform only depends on $(u,\theta,\phi)$. When we consider a source located luminosity distance $D$ away, the waveform depends on $(u,D,\theta,\phi)$ and the dependence on $D$ is proportional to $\frac{1}{D}$. In GW data analysis community, people use $t=u$ to denote the time. So we choose to use notation `$t$' for the Bondi time in the current paper to avoid two different notations for the same quantity. In order to borrow the well known relations in BMS theory, we use the Newmann-Penrose formalism and the tetrad choice convention of \cite{PenRin88,Held1970}
\begin{align}
n^0=\frac{1}{2\alpha},&\,\, n^i=-\frac{\beta^i}{2\alpha}-\frac{1}{2}v^i,\\
l^0=\frac{1}{\alpha},&\,\, l^i=-\frac{\beta^i}{\alpha}+v^i,\\
m^0=0,&\,\, m^i=\frac{1}{\sqrt{2}}(w^i-iu^i),
\end{align}
where $v^i$ is the out-pointing normal vector of the BS coordinate sphere in the 3-dimensional space-like slice, $u^i$ and $w^i$ are orthnormal basis tangent to the sphere. $v^i$ also corresponds to the propagating direction of the gravitational wave. $\alpha$ and $\beta^i$ are the lapse function and shift vector describing the 3+1 decomposition. Asymptotically $\alpha\rightarrow1$, $\beta^i\rightarrow0$, $v^i\rightarrow\frac{\partial}{\partial r}$, $u^i\rightarrow\frac{1}{r\sin\theta}\frac{\partial}{\partial\phi}$ and $w^i\rightarrow\frac{1}{r}\frac{\partial}{\partial\theta}$. Note the above convention admits a factor $\sqrt{2}$ for null vectors $\textbf{l}$ and $\textbf{n}$ difference to the convention used by numerical relativity community (for an example, Eq.~(32)-(34) of \cite{PhysRevD.77.024027}).

Based on the tetrad choice given above, we have the following relations at null infinity for asymptotically flat spacetime
\begin{align}
&\dot{\Psi}_2^{\circ}=\eth\Psi_3^{\circ}+\sigma^{\circ}\Psi_4^{\circ},\,\, \Psi_3^{\circ}=-\eth\dot{\bar{\sigma}}^{\circ},\,\, \Psi_4^{\circ}=-\ddot{\bar{\sigma}}^{\circ}.\label{eq1}
\end{align}
Here $\sigma$ corresponds to the shear of the $(\theta,\phi)$ coordinate sphere in the BS coordinate \cite{he2015new,he2016asymptotical,sun2019binary}. $\Psi_2^{\circ}$ is the Weyl tensor component relating to Bondi mass. The sign ``$\circ$" means the leading order respect to the luminosity distance when one goes to null infinity. For a function $f$ with spin-weight $s$ on sphere, the operator $\eth$ is defined as
\begin{align}
\eth f\equiv\frac{1}{\sqrt{2}}(\sin\theta)^s(\frac{\partial}{\partial\theta}
+\frac{i}{\sin\theta}\frac{\partial}{\partial\phi})(\sin\theta)^{-s}f.
\end{align}

If the tetrad convention of numerical relativity community is used, the gravitational wave strain $h\equiv h_+-ih_\times$ is related to the double integral of $\Psi_4$ respect to time (Eq.~(14) of \cite{PhysRevD.75.124018}). $h_{+}$ and $h_{\times}$ correspond to the two polarization modes of the gravitational wave \cite{maggiore2008gravitational} respect to the basis
\begin{align}
&e^+_{ij}=w_iw_j-u_iu_j\\
&e^{\times}_{ij}=w_iu_j+w_ju_i.
\end{align}
Due to the factor $\sqrt{2}$ difference of $\textbf{n}$, we now have
\begin{align}
h=-\frac{1}{2}\int\int\Psi_4dtdt=-\frac{1}{2}\int\int\frac{\Psi_4^{\circ}}{D}dtdt
\end{align}
where $D$ is the luminosity distance between the observer and the source. Again we need to note that the convention of $\Psi_4$ definition we adopt here follows \cite{PenRin88,Held1970} which admits a minus sign difference to the convention used in numerical relativity (for example, Eq.~(9) of \cite{PhysRevD.75.124018}).
Aided with the third equation of Eq.~(\ref{eq1}) we have
\begin{align}
\sigma^\circ=\frac{D}{2}\left(h_++ih_\times\right).\label{meq10}
\end{align}

And more the relations (\ref{eq1}) result in
\begin{align}
&\frac{\partial}{\partial t}(\Psi_2^{\circ}+\sigma^{\circ}\dot{\bar{\sigma}}^{\circ})=\dot{\Psi}_2^{\circ}+\dot{\sigma}^{\circ}\dot{\bar{\sigma}}^{\circ}+\sigma^{\circ}\ddot{\bar{\sigma}}^{\circ}\\
&=\eth\Psi_3^{\circ}+\sigma^{\circ}\Psi_4^{\circ}+\dot{\sigma}^{\circ}\dot{\bar{\sigma}}^{\circ}+\sigma^{\circ}\ddot{\bar{\sigma}}^{\circ}\\
&=-\eth^2\dot{\bar{\sigma}}^{\circ}-\sigma^{\circ}\ddot{\bar{\sigma}}^{\circ}+\dot{\sigma}^{\circ}\dot{\bar{\sigma}}^{\circ}+\sigma^{\circ}\ddot{\bar{\sigma}}^{\circ}\\
&=|\dot{\sigma}^{\circ}|^2-\eth^2\dot{\bar{\sigma}}^{\circ},
\end{align}
which corresponds to the `final formula' of \cite{Fra92}. $\left.h_{+,\times}\right|_{t_1=-\infty}^{t_2=\infty}$ and correspondingly $\left.\sigma^\circ\right|_{t_1=-\infty}^{t_2=\infty}$ are the gravitational wave memory.

Consequently we have
\begin{align}
\int_{t_1}^{t_2}(|\dot{\sigma}^{\circ}|^2-\eth^2\dot{\bar{\sigma}}^{\circ})dt=\left.(\Psi_2^{\circ}+\sigma^{\circ}\dot{\bar{\sigma}}^{\circ})\right|_{t_1}^{t_2},
\end{align}
which only gives the relation among the asymptotic quantities of a radiative spacetime. This relation indicates that $\left.\sigma^\circ\right|_{t_1=-\infty}^{t_2=\infty}$, i.e. gravitational wave memory, generally does not vanish. But this is just a qualitative result. It does not show the quantitative behavior of memory.

In order to investigate the quantitative behavior of GW memory, we use spin-weighted $-2$ spherical harmonic functions ${}^{-2}Y_{lm}$ to decompose the gravitational wave strain $h$ as following \cite{PhysRevD.75.124018,PhysRevD.77.024027,PhysRevD.78.124011}
\begin{align}
&h(t,\theta,\phi)\equiv\sum_{l=2}^{\infty}\sum_{m=-l}^lh_{lm}(t)[{}^{-2}Y_{lm}](\theta,\phi),\\&[{}^sY_{lm}]\equiv(-1)^s\sqrt{\frac{2l+1}{4\pi}}d^l_{m(-s)}(\theta)e^{im\phi},\\
&d^l_{ms}\equiv\sum_{i=C_1}^{C_2}\frac{(-1)^i\sqrt{(l+m)!(l-m)!(l+s)!(l-s)!}}{(l+m-i)!(l-s-i)!i!(i+s-m)!}\nonumber\\
&\times[\cos(\theta/2)]^{2l+m-s-2i}[\sin(\theta/2)]^{2i+s-m},\\
&C_1=\max(0,m-s),\,\,\, C_2=\min(l+m,l-s),
\end{align}
where the over-bar means the complex conjugate.

Noting more
\begin{align}
\eth[{}^sY_{lm}]=-\frac{1}{\sqrt{2}}\sqrt{(l-s)(l+s+1)}[{}^{s+1}Y_{lm}],
\end{align}
we have
\begin{align}
&\eth^2\dot{\bar{\sigma}}^{\circ}=\frac{1}{4}\sum_{l=2}^{\infty}\sum_{m=-l}^l \dot{h}_{lm}
\sqrt{l(l-1)(l+1)(l+2)}[{}^0Y_{lm}]\\
&|\dot{\sigma}^{\circ}|^2=\dot{\bar{\sigma}}^{\circ}\dot{\sigma}^{\circ}\\
&=\frac{1}{4}\sum_{l'=2}^{\infty}\sum_{l''=2}^{\infty}
\sum_{m'=-l'}^{l'}\sum_{m''=-l''}^{l''}
\dot{h}_{l'm'}\dot{\bar{h}}_{l''m''}\nonumber\\
&\times[{}^{-2}Y_{l'm'}]\overline{[{}^{-2}Y_{l''m''}]}\\
&=\frac{1}{4}\sum_{l'=2}^{\infty}\sum_{l''=2}^{\infty}
\sum_{m'=-l'}^{l'}\sum_{m''=-l''}^{l''}
\dot{h}_{l'm'}\dot{\bar{h}}_{l''m''}\nonumber\\
&\times[{}^{-2}Y_{l'm'}](-1)^{m''}[{}^{2}Y_{l''-m''}].\label{eq2}
\end{align}
Using the following relations \cite{Held1970,Fav09a}
\begin{align}
&\overline{[{}^sY_{lm}]}=(-1)^m[{}^{-s}Y_{l-m}],\\
&[{}^{-2}Y_{l'm'}]\overline{[{}^{-2}Y_{l''m''}]}=[{}^{-2}Y_{l'm'}](-1)^{m''}[{}^{2}Y_{l''-m''}]\nonumber\\
&=\sum_{l=0}^{\infty}
\sum_{m=-l}^l(-1)^{m+m''}\Gamma^{2-20}_{l'l''lm'-m''-m}[{}^0Y_{lm}],\\
&\Gamma^{s's''s}_{l'l''lm'm''m}\equiv\int[{}^{-s'}Y_{l'm'}][{}^{-s''}Y_{l''m''}][{}^{-s}Y_{lm}]\sin\theta d\theta d\phi,\label{eq6}\\
&\Gamma^{2-20}_{l'l''lm'-m''-m}\equiv(-1)^{m+m''}\int[{}^{-2}Y_{l'm'}]\overline{[{}^{-2}Y_{l''m''}]}\nonumber\\
&\times\overline{[{}^{0}Y_{lm}]}\sin\theta d\theta d\phi.
\end{align}
we can reduce Eq.~(\ref{eq2}) more
\begin{align}
|\dot{\sigma}^{\circ}|^2&=\frac{1}{4}\sum_{l=0}^{\infty}\sum_{m=-l}^l\sum_{l'=2}^{\infty}\sum_{l''=2}^{\infty}
\sum_{m'=-l'}^{l'}\sum_{m''=-l''}^{l''}(-1)^{m+m''}\nonumber\\
&\times\dot{h}_{l'm'}\dot{\bar{h}}_{l''m''}\Gamma^{2-20}_{l'l''lm'-m''-m}[{}^0Y_{lm}].
\end{align}
Eq.~(\ref{eq1}) reduces to
\begin{align}
&\sum_{l'=2}^{\infty}\sum_{l''=2}^{\infty}\sum_{m'=-l'}^{l'}\sum_{m''=-l''}^{l''}
\left(\int_{t_1}^{t_2}\dot{h}_{l'm'}\dot{\bar{h}}_{l''m''}dt-\right.\nonumber\\
&\left.\dot{h}_{l'm'}(t_2)\bar{h}_{l''m''}(t_2)+\dot{h}_{l'm'}(t_1)\bar{h}_{l''m''}(t_1)\right)(-1)^{m''+m}\times\nonumber\\
&\Gamma^{2-20}_{l'l''lm'-m''-m}-\sqrt{\frac{(l+2)!}{(l-2)!}}
h_{lm}\bigg{|}_{t_1}^{t_2}=4R_{lm}\bigg{|}_{t_1}^{t_2}\label{eq3}\\
&R_{lm}(t)\equiv\int \Psi_2^{\circ}(t,\theta,\phi)[{}^0Y_{lm}]\sin\theta d\theta d\phi
\end{align}
for any $l=0,1,...$ and $m=-l,...,l$. In order to unify the form of Eq.~(\ref{eq3}) we have introduced the notations $h_{00}=h_{10}=h_{1\pm1}=0$. For $l\geq2$ Eq.~(\ref{eq3}) can also be written as
\begin{align}
&h_{lm}\bigg{|}_{t_1}^{t_2}=-\sqrt{\frac{(l-2)!}{(l+2)!}}\left[\left.\frac{4}{D}\int\Psi_2^{\circ}[{}^0Y_{lm}]\sin\theta d\theta d\phi\right|_{t_1}^{t_2}\right.-\nonumber\\
&\,\,\,\,D\sum_{l'=2}^{\infty}\sum_{l''=2}^{\infty}\sum_{m'=-l'}^{l'}\sum_{m''=-l''}^{l''}
\Gamma_{l'l''lm'-m''-m}\times\nonumber\\
&\,\,\,\,\,\left(\int_{t_1}^{t_2}\dot{h}_{l'm'}\dot{\bar{h}}_{l''m''}dt-\dot{h}_{l'm'}(t_2)\bar{h}_{l''m''}(t_2)+\right.\nonumber\\
&\,\,\,\,\,\,\,\,\,\,\,\,\,\,\,\,\,\,\,\,\,\,\,\,\,\,\,\left.\left.\dot{h}_{l'm'}(t_1)\bar{h}_{l''m''}(t_1)\right)\right],\label{meq4}
\end{align}
This is a set of coupled equations for unknowns $h_{l0}$ respect to $m\neq0$ modes $h_{lm}$. For non-precession binary black holes, the gravitational wave memory is dominated by modes $h_{l0}$. Correspondingly we call $h_{l0}$ GW memory modes while $h_{lm},\,m\neq0$ non-memory modes. But for precession binary black holes this is not true anymore \cite{PhysRevD.98.064031}. Consequently we consider only spin-aligned binary black holes in the current paper. The unknowns $h_{l0}$ appear on both left and right hand sides. It is hard to solve these unknowns directly.

Due to the quasi-direct current (DC) behavior of the gravitational wave memory \cite{Fav09a}, $\dot{h}_{l0}\approx0$, and we get
\begin{align}
&h_{l0}\bigg{|}_{t_1}^{t_2}=-\sqrt{\frac{(l-2)!}{(l+2)!}}\left[\left.\frac{4}{D}\int \Psi_2^{\circ}[{}^0Y_{l0}]\sin\theta d\theta d\phi\right|_{t_1}^{t_2}\right.-\nonumber\\
&D\sum_{l'=2}^{\infty}\sum_{l''=2}^{\infty}\sum_{\mbox{\tiny$\begin{array}{c}
m'=-l',\\
m'\neq0\end{array}$}}^{l'}\sum_{\mbox{\tiny$\begin{array}{c}
m''=-l'',\\
m''\neq0\end{array}$}}^{l''}
\Gamma_{l'l''lm'-m''0}\times\nonumber\\
&\,\,\,\,\,\left(\int_{t_1}^{t_2}\dot{h}_{l'm'}\dot{\bar{h}}_{l''m''}dt-\dot{h}_{l'm'}(t_2)\bar{h}_{l''m''}(t_2)+\right.\nonumber\\
&\,\,\,\,\,\,\,\,\,\,\,\,\,\,\,\,\,\,\,\,\,\,\,\,\,\,\,\left.\left.\dot{h}_{l'm'}(t_1)\bar{h}_{l''m''}(t_1)\right)\right].\label{meq2}
\end{align}

At the past infinity time, if we take the mass center frame of the whole system as the asymptotic inertial frame, we have $\Psi_2^\circ(-\infty,\theta,\phi)=M_0$. Here $M_0$ corresponds to the Bondi mass at the past infinity time which equals to the system's ADM mass also \cite{Ashtekar:2019viz}. At the future infinity time, the Bondi mass $M$ is smaller than the initial value $M_0$ because the gravitational wave carries out some energy $E_{\rm GW}$, $M=M_0-E_{\rm GW}$. The spacetime will settle down to a Kerr black hole with mass $\tilde{M}$ at the future infinity time. But importantly the mass center frame at the future infinity time is different to the mass center frame at the past infinity time due to the kick velocity. These two asymptotic inertial frames are related by a boost transformation. Consequently $\tilde{M}=M/\gamma$, where $\gamma$ is the Lorentz factor. It is useful to note that there is not an asymptotic inertial frame which coincides with the mass center frame at all instant time due to the kick velocity. The gravitational waveform calculated by numerical relativity corresponds to the asymptotic inertial frame which corresponds to the initial mass center frame. Consequently the waveform got by numerical relativity already counts the kick velocity effect \cite{Varma:2020nbm,PhysRevLett.121.191102,PhysRevLett.117.011101}. So if we take the mass center frame at the past infinity time as the asymptotic inertial frame, we have \cite{Ashtekar:2019viz}
\begin{align}
&\Psi_2^\circ(\infty,\theta,\phi)=-\frac{\tilde{M}}{\gamma^3}\times\nonumber\\
&\left(1-v_x\sin\theta\cos\phi-v_y\sin\theta\sin\phi-v_z\cos\theta\right)^{-3},\label{meq3}\\
&\gamma=\frac{1}{\sqrt{1-v^2}},
\end{align}
where $\vec{v}$ is the kick velocity.

Since both the gravitational wave energy $E_{\rm GW}$ and the kick velocity can be calculated through non-memory modes, the right hand side of the Eq.~(\ref{meq2}) is completely determined by the non-memory modes $h_{lm},m\neq0$. If only the non-memory modes are known, we can plug them into the Eq.~(\ref{meq2}) and calculate the memory modes exactly.

If we neglect the kick velocity to let $\vec{v}=0$ and neglect the contribution $\left.\dot{h}_{l'm'}\bar{h}_{l''m''}\right|_{t_1}^{t_2}$, our Eq.~(\ref{meq2}) recovers the assumption relation (\ref{assumeq}) (equivalently, the Eq.~(3.3) of \cite{Fav09a}). The $\Psi_2^\circ$ term is understood as ordinary memory part in \cite{2020arXiv200906351K}.
\section{Comparison to previous results}
\begin{figure}
\begin{tabular}{c}
\includegraphics[width=0.5\textwidth]{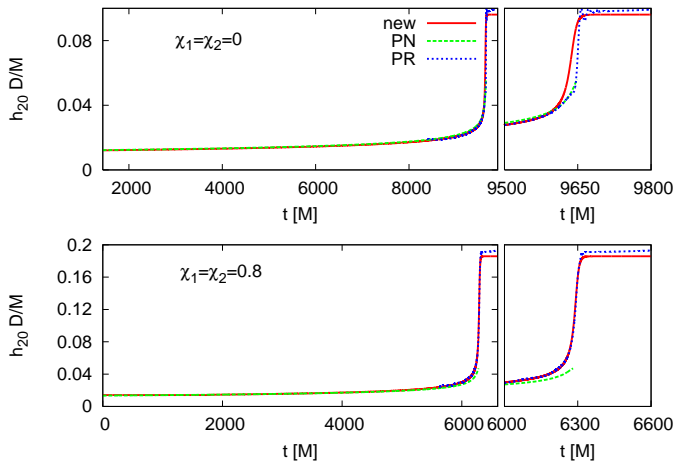}
\end{tabular}
\caption{Comparison of $h_{20}$ waveforms among the current calculation result (marked with `new'), the post-Newtonian result (marked with `PN') and the numerical relativity result (marked with `PR') for binary black hole coalescence. The top row corresponds to spinless equal mass binary black hole. The bottom row corresponds to the spin aligned equal mass binary black hole with dimensionless spin parameter $\chi_{1}=\chi_{2}=0.8$. The right two plots are the enlargement of the merger part corresponding to the left two plots.}\label{mfig1}
\end{figure}

\begin{figure}
\begin{tabular}{c}
\includegraphics[width=0.5\textwidth]{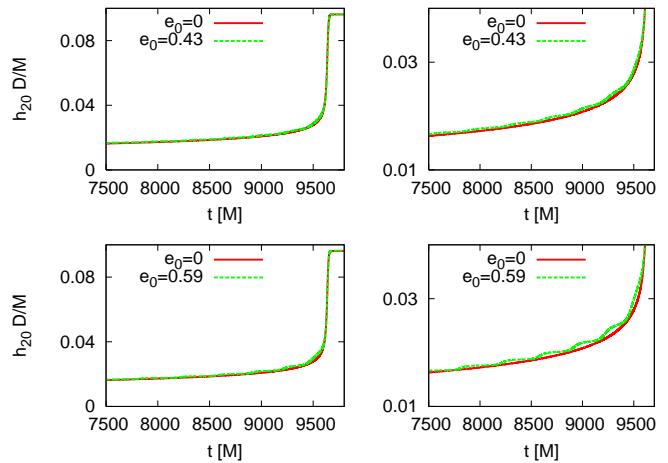}
\end{tabular}
\caption{Memory waveform for eccentric binary black holes. The eccentricity $e_0$ means the initial eccentricity at reference frequency $Mf_0=0.002$ which is got by SEOBNRE model \cite{PhysRevD.101.044049}. Here the quasi-circular one marked with $e_0=0$ corresponds to SXS:BBH:0070, the eccentric one marked with $e_0=0.43$ corresponds to SXS:BBH:1357 and the eccentric one marked with $e_0=0.59$ corresponds to SXS:BBH:1362. The right column shows
the enlargement of the inspiral part of the waveform corresponding to the left column.
}\label{mfig2}
\end{figure}

In the following we plug the SXS catalog results \cite{SXSBBH} for non-memory modes $h_{lm},m\neq0$ into the Eq.~(\ref{meq2}) to calculate memory waveform $h_{l0}$. More specifically all $l=2,..,8$ with $m=-l,...,-1,1,...,l$ modes are used. As an example we show the comparison of the $h_{20}$ calculated through the above method to the numerical relativity result \cite{PolRei11} and post-Newtonian result \cite{WisWil91,BlaDam92,Tho92,Fav09a,Fav09b,favata2009gravitational,favata2010gravitational,PhysRevD.84.124013,PhysRevD.95.084048,Cao16} in the Fig.~\ref{mfig1}. Here the post-Newtonian result is got through the Eq.~(8) of \cite{Cao16} based on the SEOBNRE model \cite{PhysRevD.96.044028,PhysRevD.101.044049}. The line corresponding to the post-Newtonian stops when the binary merger starts.

The Fig.~\ref{mfig1} indicates that the result based on the current new method is consistent to the PN waveform quite well. But the deviation shows up near merger. The new result is also consistent to the numerical relativity result quite well. But at the time numerical relativity simulation starts (where the line marked with `PR', Pollney and Reisswig, begins), the deviation between the PN waveform and new result is already clear. So the results \cite{PolRei11,Cao16} through attaching the PN approximation to numerical relativity simulation may admit systematic error. In general, the new result and the PR result are quantitatively consistent.

The authors of \cite{PolRei11} have computed more than ten binary black hole systems with equal mass and aligned spin. We confirm that all of those results are consistent to our calculation in the current work similar to the Fig.~\ref{mfig1}.

Favata found strong effect of eccentricity on the memory waveform in \cite{PhysRevD.84.124013} which results in oscillation of $h_{20}$. We confirm this result in the Fig.~\ref{mfig2}. But if we consider the GW memory amplitude $h^{tot}_{20}\equiv \left.h_{20}\right|_{-\infty}^{\infty}$ for binary black hole coalescence, the eccentricity effect is ignorable for almost equal mass binary black hole systems with eccentricity $e_0<0.6$ at reference frequency $Mf_0=0.002$ \cite{PhysRevD.96.044028,PhysRevD.101.044049}.

\begin{figure}
\begin{tabular}{c}
\includegraphics[width=0.5\textwidth]{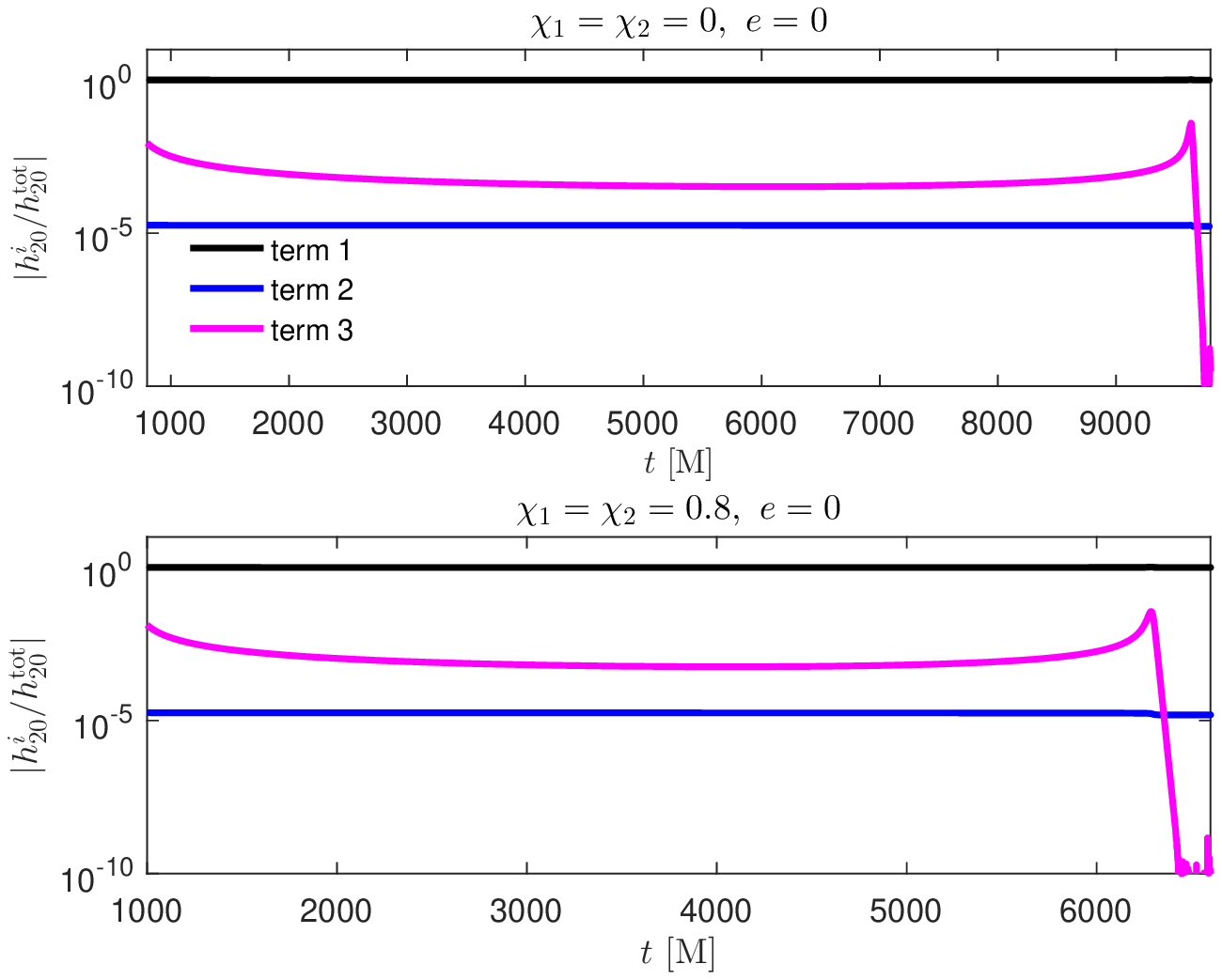}\\
\\
\includegraphics[width=0.5\textwidth]{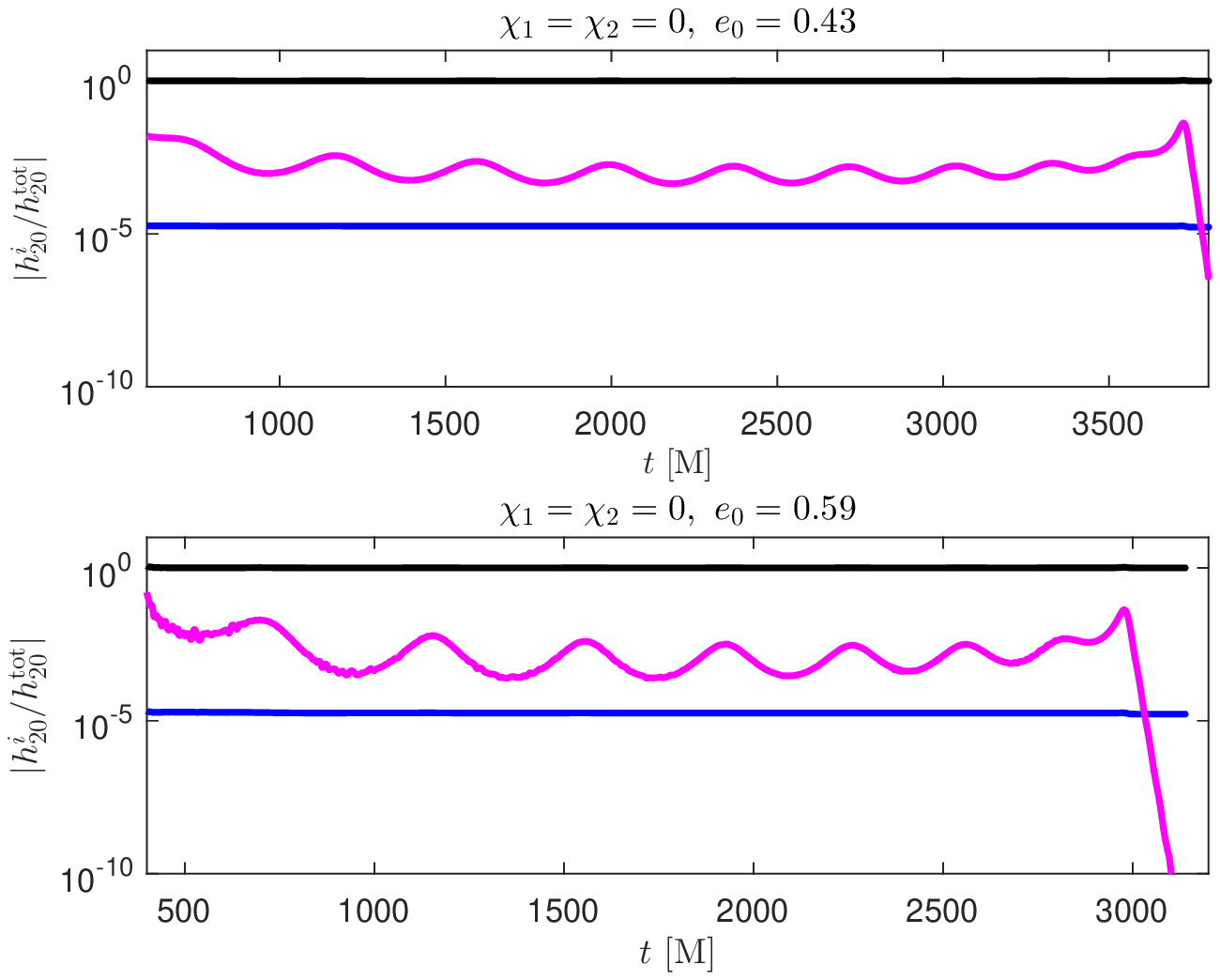}
\end{tabular}
\caption{The fractional contributions of the three terms listed in Eqs.~(\ref{term1})-(\ref{term3}) for the four cases shown in Figs.~\ref{mfig1} and \ref{mfig2}.}\label{mfig5}
\end{figure}

The assumption (\ref{assumeq}) corresponds to the term
\begin{align}
&h^1_{l0}\bigg{|}_{t_1}^{t_2}=\sqrt{\frac{(l-2)!}{(l+2)!}}D\sum_{l'=2}^{\infty}\sum_{l''=2}^{\infty}\sum_{\mbox{\tiny$\begin{array}{c}
m'=-l',\\
m'\neq0\end{array}$}}^{l'}\sum_{\mbox{\tiny$\begin{array}{c}
m''=-l'',\\
m''\neq0\end{array}$}}^{l''}\nonumber\\
&\Gamma_{l'l''lm'-m''0}\int_{t_1}^{t_2}\dot{h}_{l'm'}\dot{\bar{h}}_{l''m''}dt\label{term1}
\end{align}
of (\ref{meq2}). We call the above term 1 and denote it as $h^1_{l0}$. The authors of \cite{2020arXiv200906351K} considered the ``linear" (or alternately ``ordinary") memory contribution which corresponds to the term
\begin{align}
&h^2_{l0}\bigg{|}_{t_1}^{t_2}=-\sqrt{\frac{(l-2)!}{(l+2)!}}\frac{4}{D}\left.\int \Psi_2^{\circ}[{}^0Y_{l0}]\sin\theta d\theta d\phi\right|_{t_1}^{t_2}\label{term2}
\end{align}
of (\ref{meq2}). We call the above term 2 and denote it as $h^2_{l0}$. In addition our Eq.~(\ref{meq2}) includes instant contribution of $\dot{h}_{lm},m\neq0$
\begin{align}
&h^3_{lm}\bigg{|}_{t_1}^{t_2}=\sqrt{\frac{(l-2)!}{(l+2)!}}\sum_{l'=2}^{\infty}\sum_{l''=2}^{\infty}\sum_{m'=-l'}^{l'}\sum_{m''=-l''}^{l''}
\Gamma_{l'l''lm'-m''-m}\nonumber\\
&\times D\left(\dot{h}_{l'm'}(t_2)\bar{h}_{l''m''}(t_2)-\dot{h}_{l'm'}(t_1)\bar{h}_{l''m''}(t_1)\right).\label{term3}
\end{align}
We call the above term 3 and denote it as $h^3_{l0}$. We investigate the fractional contributions of these three terms respectively in the Fig.~\ref{mfig5} for the four cases shown in the Fig.~\ref{mfig1} and the Fig.~\ref{mfig2}. The ``linear" memory (term 2) is always negligible. The term 3 contributes between 0.01\% and 1\%. And as expected when $t\rightarrow\infty$ the term 3 vanishes. Consequently the term 3 does not contribute to $h_{lm}^{\rm tot}$. These kind of behaviors for term 2 and term 3 are common for all binary black holes.
\section{GW memory for binary black hole coalescence}
\begin{figure*}
\begin{tabular}{cc}
\includegraphics[width=0.5\textwidth]{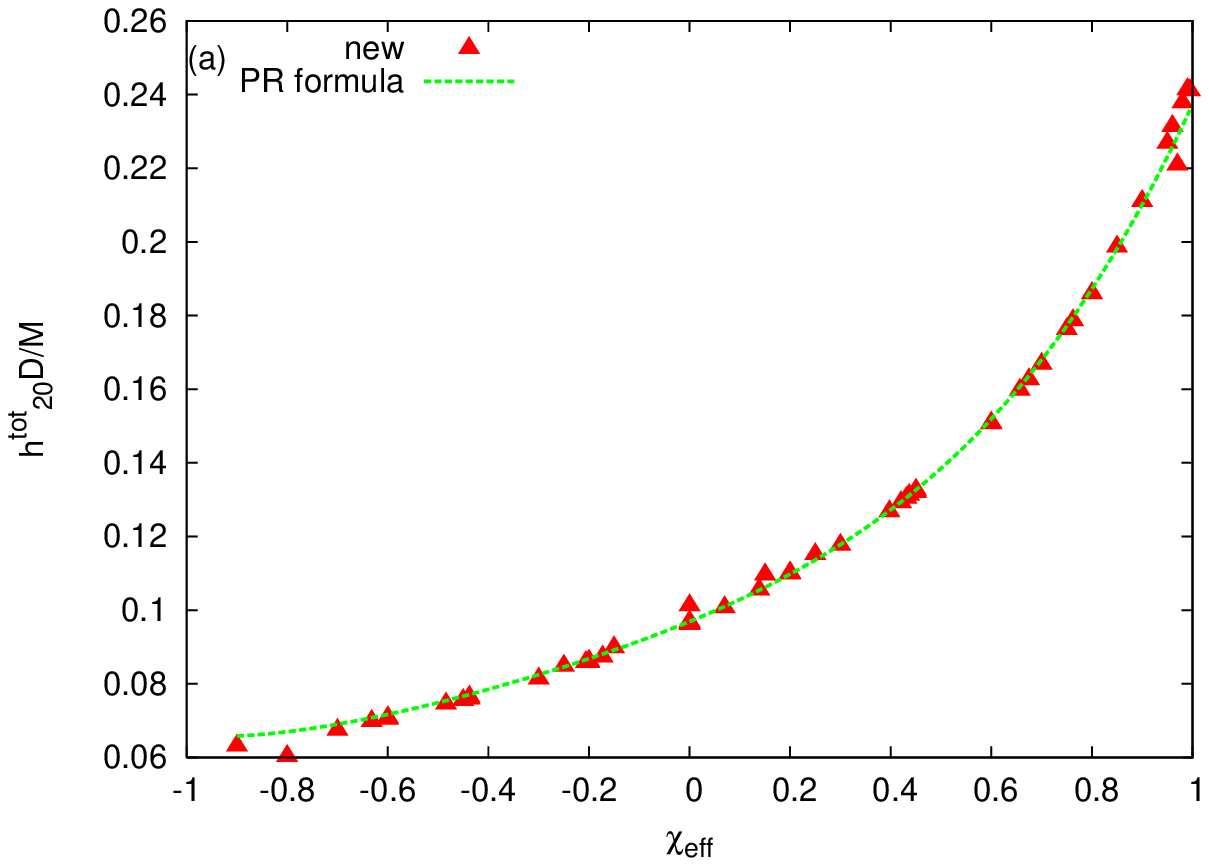}&
\includegraphics[width=0.5\textwidth]{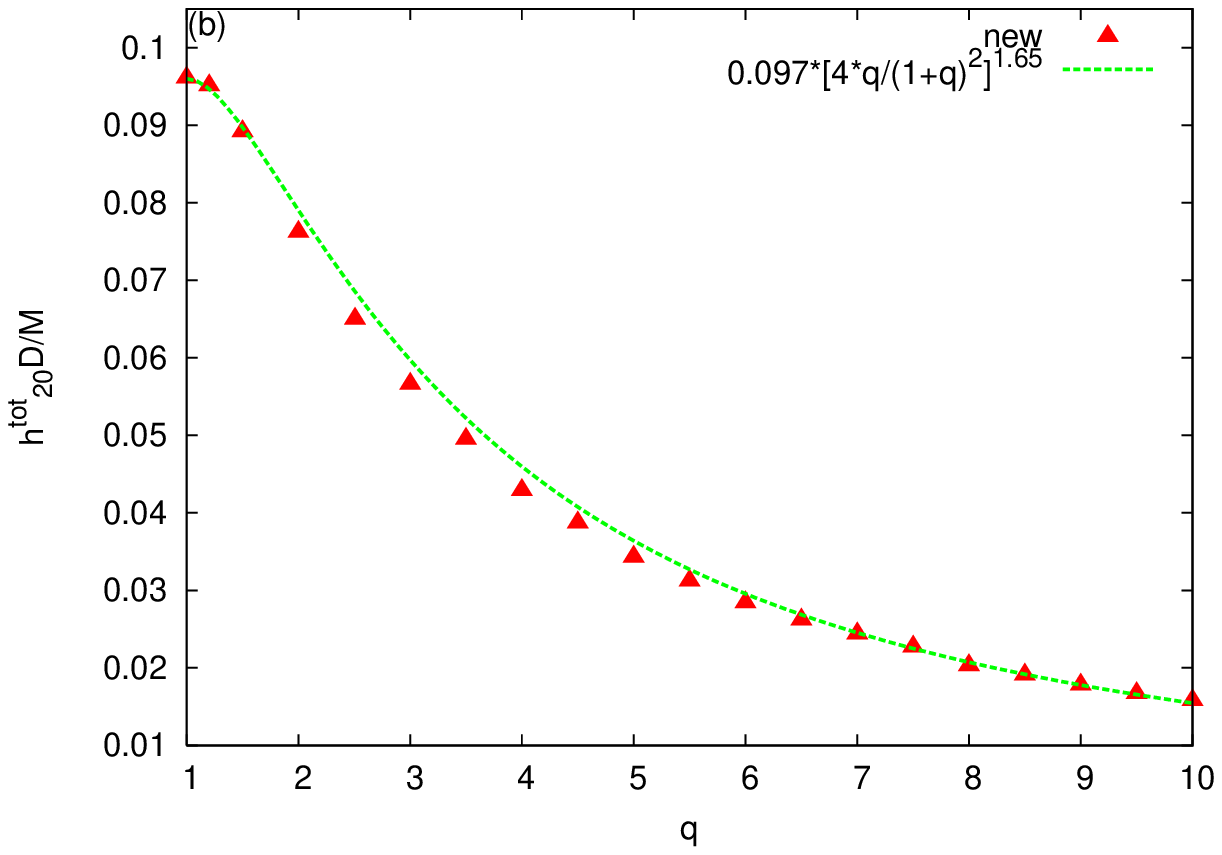}
\end{tabular}
\caption{(a) Memory amplitude $h^{tot}_{20}$ of spin aligned equal mass binary black hole respect to the effective spin $\chi_{\rm eff}\equiv\frac{q\chi_{1z}+\chi_{2z}}{1+q}$. PR formula means the Eq.~(\ref{meq5}) (also the Eq.~(8) of \cite{PolRei11}). (b) Memory amplitude of spinless binary black hole mergers respect to the mass ratio $q\equiv m_1/m_2$.}\label{mfig3}
\end{figure*}
For equal mass spin aligned binary black hole systems, the authors of \cite{PolRei11} have found the relation between the GW memory amplitude $h^{tot}_{20}$ and the symmetric spin $\chi_{\rm eff}\equiv(m_1\chi_{1z}+m_2\chi_{2z})/M$ (Eq.~(8) of \cite{PolRei11})
\begin{align}
\frac{D}{M}h^{tot}_{20}=&0.0969+0.0562\chi_{\rm eff}+0.0340\chi_{\rm eff}^2+\nonumber\\
&0.0296\chi_{\rm eff}^3+0.0206\chi_{\rm eff}^4.\label{meq5}
\end{align}
We confirm this formula based on SXS catalog in the Fig.~\ref{mfig3}(a). As pointed out by the authors of \cite{PolRei11} and also explained in \cite{Cao16}, we also confirm the memory amplitude $h^{tot}_{20}$ for equal mass spin aligned binary black hole is independent of the anti-symmetric part of the spin $\chi_{\rm A}\equiv\frac{\chi_{1z}-\chi_{2z}}{2}$.

In the Fig.~\ref{mfig3}(b), we investigate the GW memory amplitude of spinless binary black hole mergers respect to the mass ratio $q\equiv m_1/m_2$. Based on the PN approximation, Favata \cite{Set09,Fav09b} showed the memory of the binary black hole with equal mass is about $\frac{1}{24\pi^2}\sqrt{\frac{1543}{70}}\approx0.0198$ which is much less than the calculation result 0.097 here. And more Favata \cite{Set09,Fav09b} estimated the memory is proportional to the symmetric mass ratio $\eta=\frac{q}{(1+q)^2}$. Here we find that it decreases much faster than Favata estimated. Instead it roughly behaves as $\eta^{1.65}$.

\begin{figure*}
\begin{tabular}{cc}
\includegraphics[width=0.5\textwidth]{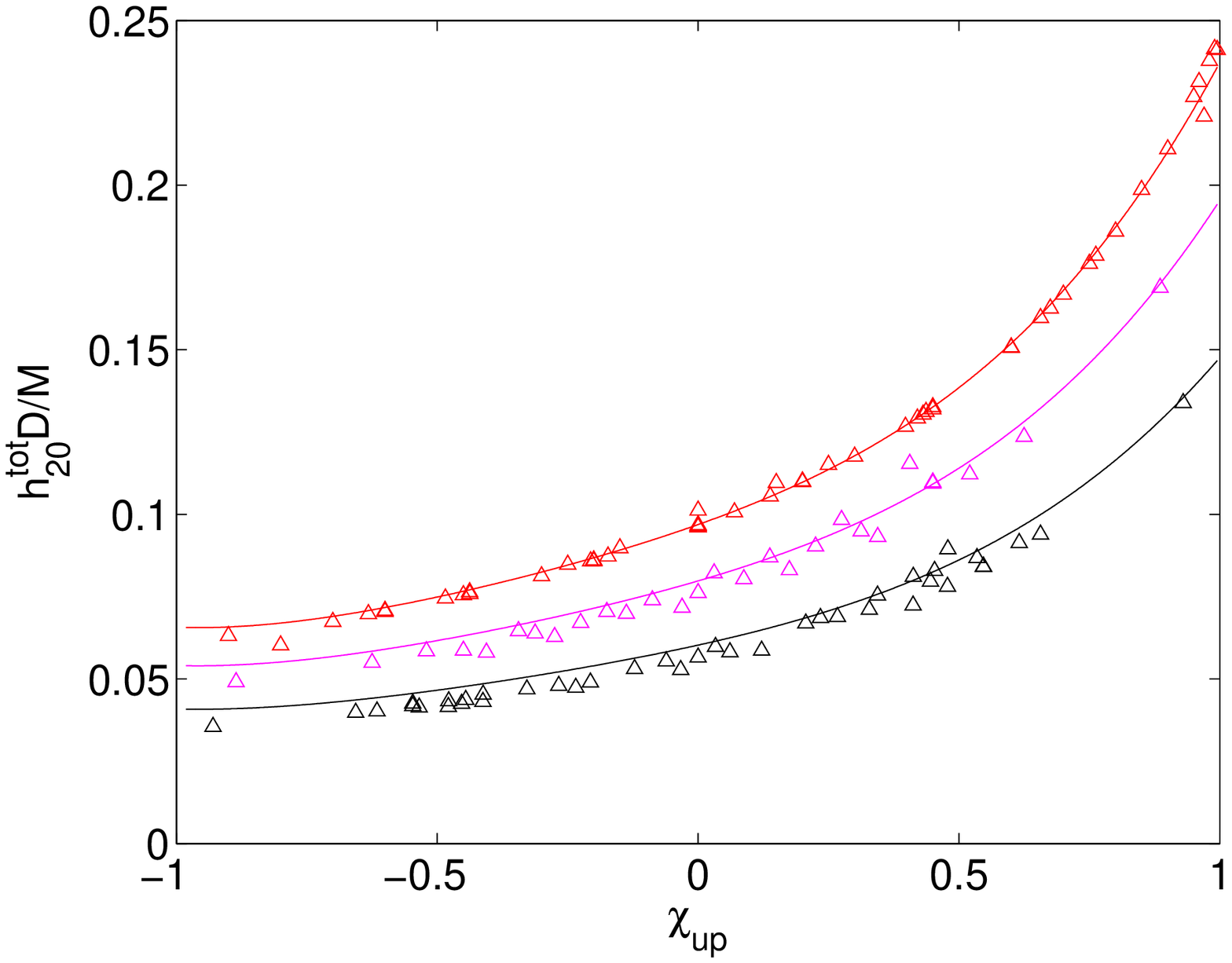}&
\includegraphics[width=0.5\textwidth]{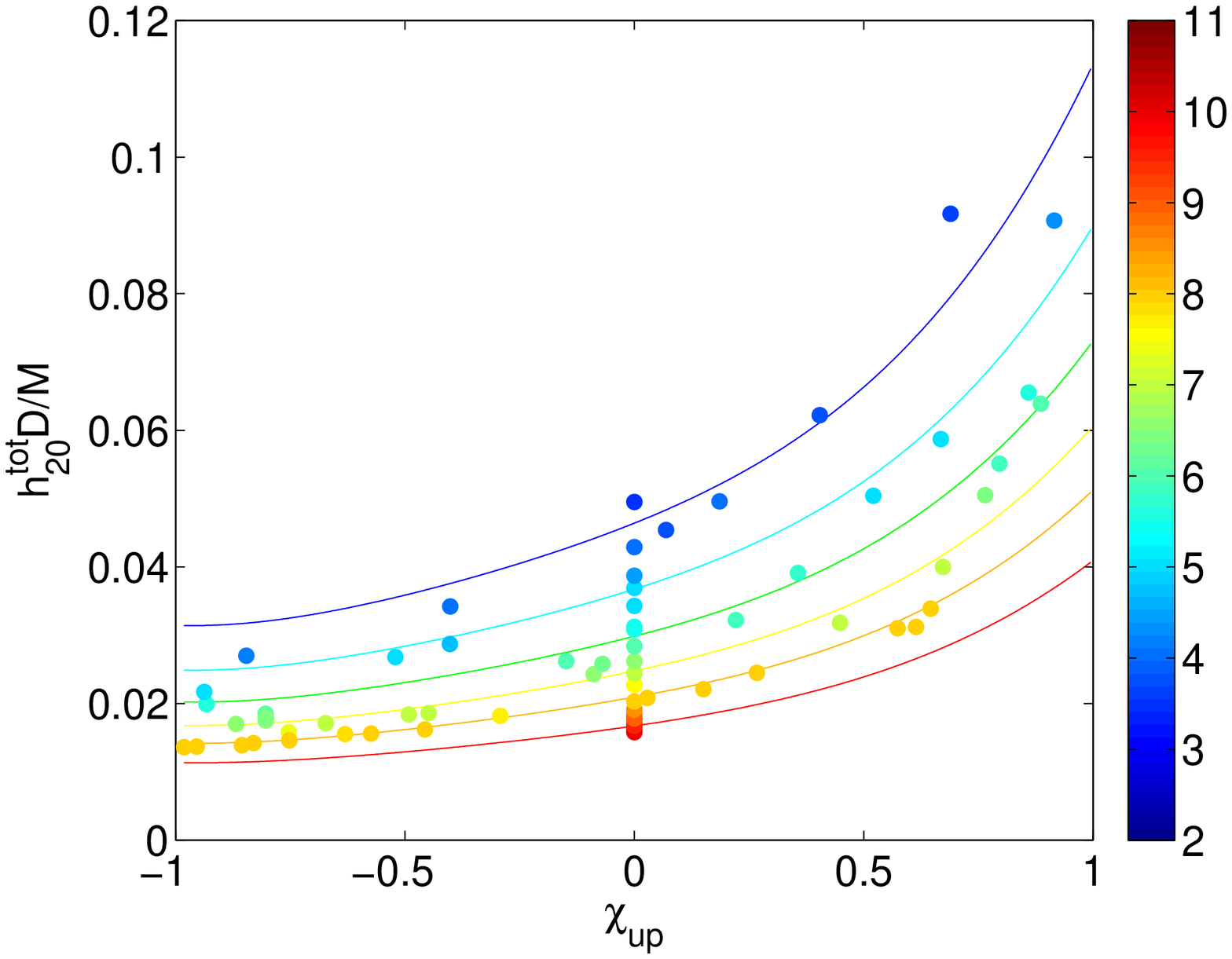}
\end{tabular}
\caption{Memory amplitude $h^{tot}_{20}$ for generic spin aligned binary black hole respect to the spin hang-up parameter $\chi_{\rm up}$ which is defined in (\ref{meq7}). The different colors mean different mass ratio. The left plot: red, cyan and black represent respectively $q=1$, $q=2$ and $q=3$. Right plot: The different lines from top to bottom represent the Eq.~(\ref{meq6}) with $q=4$, $q=5$, $q=6$, $q=7$, $q=8$ and $q=9.5$ respectively. The color of the points represent the different mass ratio indicated by the color bar.}\label{mfig4}
\end{figure*}
For unequal mass spin aligned binary black holes, the Eq.~(\ref{meq5}) does not hold any more. In \cite{PhysRevD.101.044049} we found spin hang-up effect is the most important factor for gravitational waveform. Interestingly we find this statement is also correct for memory. Following \cite{PhysRevD.101.044049} we define a spin hang-up parameter as
\begin{align}
\chi_{\rm up}\equiv\chi_{\rm eff}+\frac{3}{8}\sqrt{1-4\eta}\chi_{\rm A}.\label{meq7}
\end{align}
This definition is different to the Eq.~(7) of \cite{PhysRevD.101.044049}. The current definition lets $\chi_{\rm up}$ go back to $\chi_{\rm eff}$ for equal mass binary black holes. Based on this spin hang-up parameter and the relationship between the GW memory amplitude and the mass ratio, we find the general behavior for generic spin aligned binary black hole systems can be expressed as
\begin{align}
\frac{D}{M}h^{tot}_{20}=&[0.0969+0.0562\chi_{\rm up}+0.0340\chi_{\rm up}^2+\nonumber\\
&0.0296\chi_{\rm up}^3+0.0206\chi_{\rm up}^4](4\eta)^{1.65}.\label{meq6}
\end{align}
We validate the finding (\ref{meq6}) in the Fig.~\ref{mfig4}. From this figure we can see the Eq.~(\ref{meq6}) does describe the main feature of the behavior. For systems with mass ratio between 2 and 4, the effect of $\chi_{\rm A}$ is stronger. So the points do not perfectly fall on the line. We suspect this is because only the combination of $\chi_{\rm A}$, $\sqrt{1-4\eta}$ and $\eta$ contributes to memory for $\chi_{\rm A}$ like the Eqs.~(\ref{meq7}) and (\ref{meq6}). For the rest cases the Eq.~(\ref{meq6}) works very well. When the mass ratio increases, the effect of mass ratio decreases which can be seen in the Eq.~(\ref{meq6}). So after $q=3$ we use gradually larger and larger range to group numerical data in the Fig.~\ref{mfig4}.
\section{Discussion}
We have proposed a new method to accurately calculate the GW memory for spin-aligned binary black holes. Our calculation indicates that the strongest GW memory amplitude for binary black hole merger corresponds to the fastest spinning aligned two equal mass black holes. And the amplitude is about $h^{tot}_{20}\approx0.24\frac{M}{D}$. If the two black holes do not spin, the amplitude is about $h^{tot}_{20}\approx0.1\frac{M}{D}$. Quantitatively we find that the memory amplitude can be described by spin hang-up parameter $\chi_{\rm up}$ and mass ratio $\eta$ quite well.

Based on our new method, it is straight forward to apply the technique of \cite{PhysRevResearch.1.033015} to construct a highly accurate numerical relativity surrogate model for GW memory. In the near future the detection of GW memory can be compared to the prediction by our method \cite{2020arXiv200906351K} and give a test of general relativity.

\section*{Acknowledgments}
We thank Zhi-Chao Zhao for helpful discussions. This work was supported by the NSFC (No.~11690023). X. He was supported by NSF of Hunan province (2018JJ2073). Z. Cao was supported by ``the Fundamental Research Funds for the Central Universities", ``the Interdiscipline Research Funds of Beijing Normal University" and the Strategic Priority Research Program of the Chinese Academy of Sciences, grant No. XDB23040100.
\bibliography{refs}

\end{document}